\documentclass[aps,prl,twocolumn,showpacs,amsmath,floatfix]{revtex4}
\usepackage{amsmath,graphics,epsfig,color,verbatim,ulem}

\begin{document}

\title{Coupled magnetic-ferroelectric metal-insulator transitions in epitaxially-strained SrCoO$_{3}$ from first principles}

\author{Jun Hee Lee}
\email{jhlee@physics.rutgers.edu}
\author{Karin M. Rabe}
\affiliation{Department of Physics and Astronomy,
Rutgers University, Piscataway, New Jersey 08854-8019, USA}

\marginparwidth 2.7in
\marginparsep 0.5in

\begin{abstract}

First-principles calculations of the epitaxial-strain phase diagram of perovskite SrCoO$_{3}$ are presented. Through combination of the large spin-phonon coupling with polarization-strain coupling and coupling of the band gap to the polar distortion, both tensile and compressive epitaxial strain are seen to drive the bulk ferromagnetic-metallic (FM-M) phase to antiferromagnetic-insulating-ferroelectric (AFM-I-FE) phases, the latter having unusually low elastic energy.
At these coupled magnetic-ferroelectric metal-insulator phase boundaries, cross responses to applied electric and magnetic fields and stresses are expected. In particular, a magnetic field or compressive uniaxial stress applied to the AFM phases could induce an insulator-metal transition, and an electric field applied to the FM-M phase could induce a metal-insulator transition. 

\end{abstract}
\pacs{71.30.+h, 77.80.-e, 75.85.+t, 63.20.-e}

\maketitle

Recent dramatic advances in the synthesis of complex oxides have opened exciting avenues for the theoretical design and laboratory realization of new classes of functional oxide materials\cite{oxides}.  With tuning of a control parameter such as composition, artificial structuring or epitaxial strain, the competition of structural, magnetic and electronic degrees of freedom characteristic of complex oxides can result in two phases very close in energy but with very different structure and magnetic ordering. Desirable functional behavior can be obtained with the application of applied fields and/or stresses that produce switching between the two phases, a metal-insulator transition \cite{MIT} triggered by electric or magnetic fields or applied stress being of particular current interest. \cite{vanox1,vanox2}

As an extension of previous work on epitaxial-strain-induced multiferroicity in EuTiO$_3$ \cite{Rabe06,Darrell10} and SrMnO$_3$ \cite{ours-SMO}, one strategy to identify the parent phase for a coupled magnetic-ferroelectric metal-insulator transition is to look for a high-symmetry metallic phase with a lattice distortion that is strongly destabilized by a change in magnetic order and, moreover, opens a band gap. This would result in a low-energy alternative phase, with distinct magnetic order, a lattice distortion and a band gap, that could be stabilized by tuning of an appropriate control parameter, establishing a phase boundary at which switching between the metallic and insulating phases would be possible.

Here, we draw on the results of a first-principles survey of the spin-phonon coupling in selected sets of magnetic perovskite oxides \cite{5-compound} to identify SrCoO$_3$ as a suitable parent phase. These calculations showed that cubic SrCoO$_3$ has a large spin-phonon coupling, with the lowest-frequency polar phonon being strongly destabilized by a change in magnetic order from ferromagnetic (FM) to the higher-energy $G$-AFM ordering. In experimental observations, bulk SrCoO$_3$ is found to be cubic, with ferromagnetic order below $T_{\rm c}$=280$\sim$305 K \cite{RT-FM1,RT-FM2,RT-FM3}. Previously reported experiments on SrCoO$_3$ show that oxygen vacancies or chemical doping can lead to dramatic changes in structure, magnetic ordering and resistivity \cite{RT-FM2,AF1,AF2,MR1,MR2,MR3,MR4,MR5,transition1}. The large spin-phonon coupling seen in first-principles results suggests that similarly dramatic coupled structural, magnetic and metal-insulator phase transitions, and associated functional properties, could occur with epitaxial strain or isoelectronic substitution.

In this paper, we investigate the ground-state epitaxial strain phase diagram of SrCoO$_3$ from first principles and find phase transitions from the bulk FM-M phase to insulating AFM-FE phases occurring at accessible values of compressive and tensile epitaxial strain; furthermore, the energies of the strained phases are exceptionally low with respect to the bulk, promoting their experimental realization. We determine the magnetic ordering, the metallic (M) or insulating (I) character, and the polarization of ferroelectric (FE) phases as a function of epitaxial strain. We find that it is the polar distortion that is responsible for opening the band gap in the insulating phases, which offers the possibility of epitaxial-strain and electric-field tuning of the band gap in the AFM-FE phase.
We predict large mixed magnetic-electric-elastic responses in the vicinity of the phase boundaries resulting from switching between the two phases, including metal-insulator transitions triggered by electric and magnetic fields and uniaxial stress. 

First-principles calculations were performed using density-functional 
theory within the
generalized gradient approximation GGA+$U$ method \cite{GGAU}
with the Perdew-Becke-Erzenhof parametrization \cite{PBE}
as implemented in
the $Vienna$ $Ab$ $Initio$ $Simulation$ $Package$ 
(VASP-4.6)~\cite{Kresse2,Kresse3}.  
The projector augmented wave (PAW) potentials \cite{Kresse1} explicitly
include 10 valence electrons for Sr (4$s^2$4$p^6$5$s^2$), 9 for Co
(3$d^8$4$s^1$), and 6 for oxygen (2$s^2$2$p^4$). 
The wave functions are expanded in a plane waves basis with an energy cut-off of 500 eV. 
We use the Dudarev~\cite{Dudarev} implementation
with on-site Coulomb interaction $U$=2.5 eV
and on-site exchange interaction $J_H$=1.0 eV
to treat the localized $d$ electron states in Co.
The value of $U$ chosen gives a value of 2.6$\mu_B$/f.u. and 
a lattice constant $a_0$= 3.842 \AA for the cubic FM structure in good agreement with 
a previous calculation \cite{ground,ground1} and recent measurements (2.5$\mu_B$ \cite{RT-FM3}, $a_0$=3.835 \AA \cite{RT-FM1}) 
on stoichiometric SrCoO$_3$ samples. 
This magnetic moment corresponds to an intermediate ($t_{2g}^4$$e_{2g}^1$) spin state. 

The phonon frequencies of cubic SrCoO$_3$ with $G$-AFM and FM orderings were computed 
using the frozen phonon method at the $\Gamma$, M and R points
in the irreducible Brillouin zone of the primitive perovskite unit cell, with atomic displacements of 0.001$a_0$. 
To find the minimum-energy configuration in a given space group 
determined by freezing in one or more unstable modes of the cubic 
reference structure, we moved the
atoms according to the conjugate-gradient algorithm until the residual 
Hellman-Feynman forces were 
less than 1.0meV/\AA. 
An $8\times8\times8$ Monkhorst-Pack k-point grid was used for 5-atom unit cells, while a $4\times4\times4$ grid was used for the $\sqrt 2\times \sqrt 2\times \sqrt 2$ (10-atom) supercell and $\sqrt 2\times \sqrt 2\times 2$ (20-atom) supercell \cite{correction}.
To study the effects of epitaxial strain, we performed ``strained-bulk'' 
calculations \cite{Pertsev,Oswaldo} in which calculations were performed for the 
periodic crystal with
appropriate epitaxial constraints imposed on the in-plane lattice 
parameters, with all atomic positions and the out-of-plane lattice 
constant optimized. Epitaxial 
strain is here defined relative to the computed lattice constant for the 
FM cubic perovskite structure (3.842\AA).
Ferroelectric polarizations for the relaxed structures at each strain
were computed by the Berry-phase method \cite{pol}.


\begin{table}
\caption{Selected calculated phonon frequencies (cm$^{-1}$) 
and relative total energy (meV/f.u.) 
of cubic SrCoO$_3$ at the calculated equilibrium lattice constants with FM ($a_0$=3.842~\AA) 
and $G$-AFM ($a_0$=3.885~\AA) orderings.  
The classification of the three zone-center mode eigenvectors as Slater, Last and Axe 
follows standard terminology for perovskites \cite{mode}.}
\begin{ruledtabular}
\begin{tabular}{c|ccc|c|c|c}
                          & Slater    & Last        & Axe & R$_4^+$  &  M$_3^+$& $E_{\rm tot.}$ \\                
\hline                                                                       
FM ($a_0$=3.842\AA)       & 203       & 154         & 498 & 56       &      89 &   0       \\                
$G$-AFM ($a_0$=3.842\AA)  & 176{\it i}& 155         & 508 & 63{\it i}&      24 &  258      \\                
$G$-AFM ($a_0$=3.885\AA)  & 199{\it i}& 141         & 480 & 54{\it i}&      33 &  220      \\                
\end{tabular}
\end{ruledtabular}
\label{frequency}
\end{table}

Selected phonon frequencies of cubic FM and $G$-AFM SrCoO$_3$ are given in Table~\ref{frequency}.
A dramatic change of phonon frequency with magnetic ordering 
is seen for the Slater mode, which is stable in the ground state FM cubic structure but strongly unstable with the higher-energy $G$-AFM ordering. 
The corresponding eigenvectors of the Slater mode are (Sr,Co,O$_{\parallel}$,O$_{\perp}$) = 
(0.02,0.38,-0.32,-0.61) for FM and (0.00,0.43,-0.39,-0.58) for $G$-AFM, 
showing displacement of 
the $B$-site Co cation relative to a fairly rigid oxygen octahedron network. The Axe mode is much less sensitive to the magnetic ordering, 
while the change in the Last mode is negligible. 
In addition, the oxygen octahedron rotation modes at R and M soften considerably in the $G$-AFM state. 

\begin{figure}
\begin{center}
\includegraphics[height=12.9cm,trim=0mm 0mm 0mm 0mm]{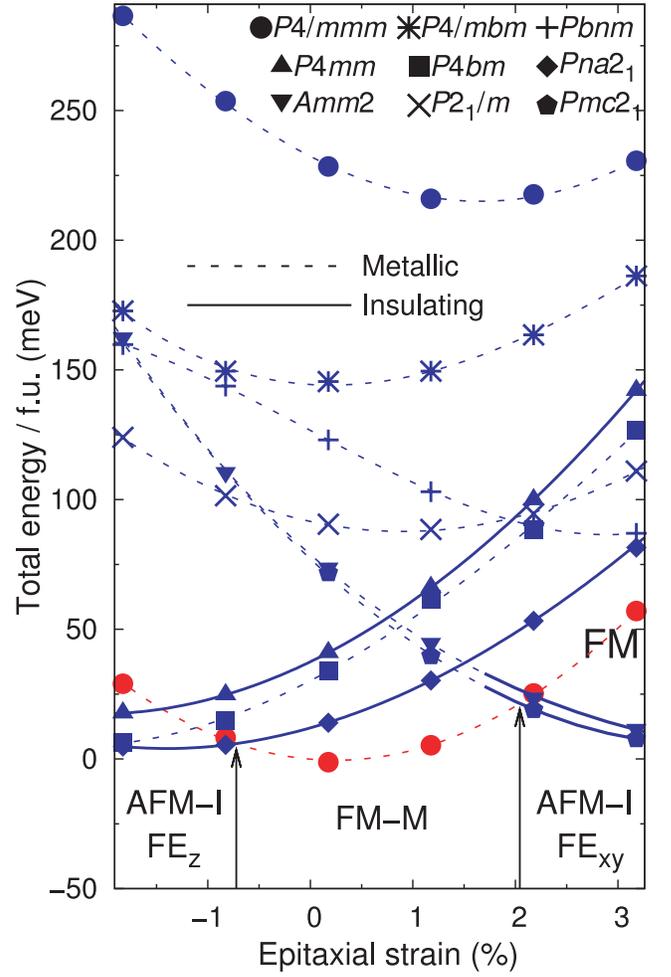}
\end{center}
\caption{(Color online) GGA+$U$
total energies of various structures labeled by space groups generated by freezing in selected modes of the primitive perovskite structure \cite{Stokes}, as listed below.
Calculations were performed at every 1\% strain increment and interpolated.
The energies of structures with FM and $G$-AFM ordering are shown with red and blue symbols,respectively; the single FM curve is also labeled.
Vertical black arrows at -0.72\% and +2.04\% strain 
indicate phase boundaries separating the $Pna2_1$ AFM-I-FE$_z$ phase, the $P$4/$mmm$ FM-M phase, and the $Pna2_1$ AFM-I-FE$_z$ phase, where 
FE$_z$ and FE$_{xy}$ denote the direction of ferroelectric polarization. $P4mm=\Gamma_4^-[001]$, $Amm2=\Gamma_4^-[110]$, $P4/mbm=M_2^+[001]$, $P4bm=M_2^+[001]+\Gamma_4^-[001]$, 
$P2_1/m=R_4^+[011]+M_3^+[100]$, $Pbnm=R_4^+[110]+M_3^+[001]$, $Pmc2_1=R_4^+[110]+M_3^+[001]+\Gamma_4^-[110]$,
$Pna2_1=R_4^+[110]+M_3^+[001]+\Gamma_4^-[001]$. 
}
\label{energy}
\end{figure}

For FM ordering, we find that the cubic structure is stable not only against the distortions described by single modes in Table~\ref{frequency}, but also against the distortion into the orthorhombic $Pbnm$ structure, common in perovskite oxides, generated by the combination $R_4^+[110]+M_3^+[001]$. This is
consistent with the experimental observation that bulk SrCoO$_3$ is cubic.  

In contrast, for $G$-AFM ordering, the cubic structure is unstable against distortion. In fact, the energy gains associated with several different distortions are comparable to 230meV/fu, the energy of the cubic G-AFM phase ($a_{\rm 0}$=3.885 \AA) relative to the FM ground state phase. 
($Pbnm$=+89meV/fu, $P4mm$=+34, $Amm2$=+15, $Pna{\rm 2}_{\rm 1}$($R_4^+[110]+M_3^+\Gamma_4^-[001]$)=+18, 
$Pmc{\rm 2}_{\rm 1}$($R_4^+[110]+M_3^+[001]+\Gamma_4^-[110]$)=+14)

In Fig.~\ref{energy}, the epitaxial strain dependences of the total 
energies of the nonpolar $P4/mmm$ structure and structures with various polar, Jahn-Teller (JT), and rotational distortions, 
are shown for FM and $G$-AFM ordering for 
strains from -1.9\% (compressive) to +3.3\% (tensile) strain. 
At 0\% strain, the ground state cubic FM phase is the lowest energy state, while the $G$-AFM $P4/mmm$ phase is much higher 
in energy. However, the polar instability of the latter phase (see Table~\ref{frequency}) leads to large energy gains for polar $G$-AFM phases. 
At 0\% strain, there is a gain relative to nonpolar $P{\rm 4}/mmm$ of
190 meV/f.u. with $\Gamma_4^-[001]$ distortion ($P{\rm 4}mm$) 
and 150 meV/f.u. with $\Gamma_4^-[110]$ ($Amm$2).  
The $P{\rm 4}mm$ phase and $Amm$2 phase are further favored by compressive and tensile strain, respectively, and their energies   
drop below that of the FM $P$4/$mmm$ phase at sufficiently large values of the strain. In addition to lowering the energy, polar distortion of the metallic $G$-AFM $P4/mmm$ phase tends to open a band gap, with the polar ($P{\rm 4}mm$) phase having a nonzero gap for all strains considered, and $Amm$2 having a nonzero gap for tensile strain above +1.7\%.

Rotational and JT distortions can also lower the total energies of certain phases with $G$-AFM ordering. 
However, we find that structures with only rotational ($Pbnm$, $P2_1/m$) or only M-type JT distortions ($P4/mbm$) of the high-symmetry $P{\rm 4}/mmm$ phase are not energetically competitive with the cubic FM phase. 
On the other hand, rotational instabilities in the polar AFM $P{\rm 4}mm$ phase further lower the energy and lower the symmetry to $Pna2_1$, lowering the critical strain of the phase boundary between FM and AFM from -1.5\% to -0.7\%. For AFM $Amm2$, the rotational instabilities lead to a smaller energy lowering and a symmetry lowering to $Pmc2_1$, with a small reduction in the critical strain. Finally, we computed the energies of phases with $A$- or $C$- type orderings, which appear in the computed epitaxial strain phase diagram of SrMnO$_3$\cite{ours-SMO}. In SrCoO$_3$, at each strain value these magnetic orderings prove to be higher in energy than the FM or $G$-AFM phases.

\begin{figure}
\begin{center}
\includegraphics[width=8.92cm,trim=11mm 2mm 4mm 0mm]{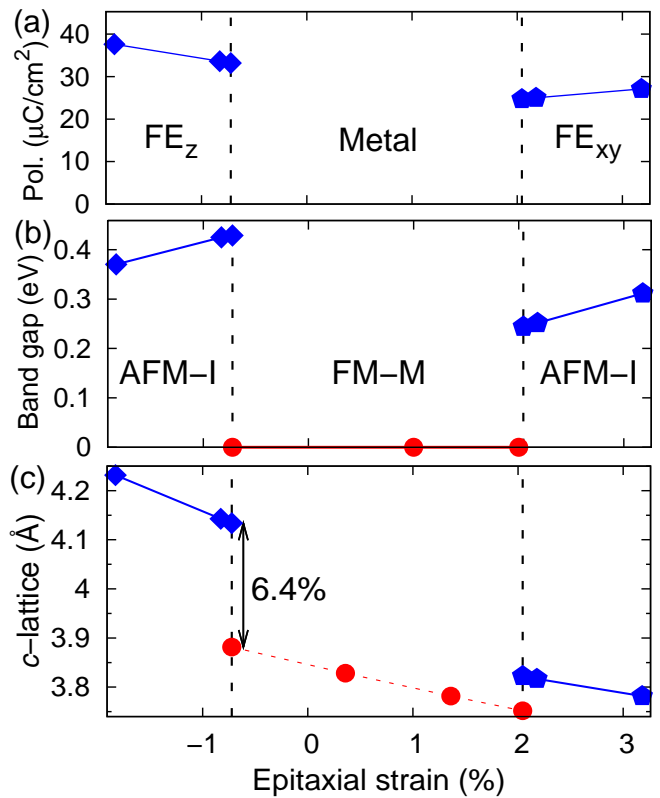}
\end{center}
\caption{(Color online) (a) Computed GGA+$U$ ferroelectric polarization 
of SrCoO$_3$ $G$-AFM (square) 
in the lowest energy structure at each strain value.
Open and solid symbols represent structures where 
ferroelectric polarization is along [110] and [001] respectively. 
(b) Band gap
(c) $c$-lattice parameter for the lowest energy structure at each strain value. Symbols as in Fig.~\ref{energy}.} 
\label{phase}
\end{figure}

Considering the lowest energy phase at each strain, we find three distinct phases in the epitaxial strain phase diagram of SrCoO$_3$, with phase boundaries at -0.72\% and +2.04\% strain. Next, we consider the computed properties of the epitaxially strained phases. 
In Fig.~\ref{phase}, we present the FE polarization, $c$-lattice parameter, and band gap of the lowest energy phase at strains from -1.9\% to +3.3\%. 
The polarizations of the FE phases are comparable to that of bulk BaTiO$_3$, and increase with increasing strain as expected from the polarization-strain coupling characteristic of ferroelectric perovskites.
The large jumps in $c$-lattice parameter from the FMM phase to the FE$_z$ phase can also be attributed to the strain-polarization coupling. 
The computed band gaps are above 0.25 eV, and these DFT values are expected to be substantial underestimates of the yet-to-be measured values.
The role of the polar distortion in opening a band gap was previously noted in the related compound SrMnO$_3$ \cite{ours-SMO}. 
While the band gap is mainly controlled by the amplitude of the polar distortion in SrCoO$_3$ as well, the rotational distortions and strain also affect the value, accounting for the epitaxial strain dependence of the gap in the FE$_z$ phase. 

Both of the phase boundaries in the computed epitaxial strain phase diagram of SrCoO$_3$ are first order, with large jumps in properties across the boundary. Thus, for a system in the vicinity of one of the phase boundaries, large mixed magnetic-electric-elastic responses 
are predicted as the result of the possibility of switching between two phases with quite different magnetic ordering, polarization, 
band gap and $c$-lattice parameter. 
In particular, a AFM-FE phase just at the phase boundary could be driven by applied magnetic field to the cubic FM phase, 
with a downward jump in $c$-lattice parameter; alternative, with applied uniaxial compressive stress, a nonzero magnetization could be induced. 
Conversely, a cubic FM phase just at the phase boundary could be driven by an appropriately oriented applied electric field 
to the adjacent AFM-FE phase, with a jump in magnetization and $c$-lattice parameter. 

The most interesting behavior associated with these phase boundaries is the possibility of a metal-insulator transition triggered by applied  electric or magnetic fields or uniaxial stress. In applied magnetic field, a AFM-FE system just at the phase boundary will not only show 
a jump in magnetization, polarization and $c$-lattice parameter, but also a sharp transition from an insulating to a metallic state. 
The jumps in $c$-lattice parameter and polarization also offer the possibility of triggering an insulator-metal transition in the AFM-FE state and a metal-insulator transition in the metallic FM state 
by applying uniaxial stress and an electric field, respectively. 
The on-off ratio is expected to be large, since the insulating phase has a band gap of at least several tenths of an eV and thus a low conductivity, while the metallic phase should have high conductivity since SrCoO$_3$ has no substitutional disorder that would increase scattering. 

The topology of the phase diagram is not sensitive to changes in U. Quantitatively, in the extreme case U=0, the computed critical strains are larger (about $\pm 4\%$) 
because the energy difference 
between FM and $G$-AFM ($\Delta E$=300 meV) is larger  
than in GGA+$U$ ($\Delta E$=258 meV), and the 
frequency shift of Slater phonon ($w$(FM)=163cm$^{-1}$ to $w$($G$-AFM)=124$i$cm$^{-1}$) 
is reduced compared to the GGA+$U$ value reported in Table~\ref{frequency}. 

In experimental investigation of these predictions, there will be two major challenges. The first, common to all systems with first-order phase transitions, will be to minimize the hysteresis in switching from one phase to another; one strategy to increase reversibility has been demonstrated in Ref. \onlinecite{James-rev}. The other, specific to SrCoO$_3$, will be to reduce the oxygen vacancy concentration to achieve stoichiometric SrCoO$_3$ epitaxial films. However, the unusually small elastic energy associated with the AFM-FE phases at and above the critical strains should greatly facilitate the growth of the strained films.

In summary, we have presented first-principles 
calculations of the epitaxial-strain phase diagram of perovskite 
SrCoO$_{3}$. Through combination of the large spin-phonon coupling with polarization-strain coupling and coupling of the band gap to the polar distortion, both tensile and compressive epitaxial strain are seen to drive the bulk FM-M phase to AFM-I-FE phases.
At these coupled magnetic-ferroelectric metal-insulator phase boundaries, cross responses to applied electric and magnetic fields and stresses are expected. 
An applied magnetic field can induce a transition from the AFM phases to the FM phase, with accompanying disappearance of polarization, downward jump in c lattice parameter, and an insulator-metal transition, the latter producing a highly nonlinear magnetoresistance. Similarly, an applied uniaxial compressive stress will change the magnetic ordering and induce an insulator-metal transition by driving the AFM-I-FE phases to the FM-M phase. Finally, an applied electric field in the FM-M cubic phase will drive the system to the AFM-I-FE phases, reducing the magnetization to zero and inducing a nonzero polarization and metal-insulator transition.

We would like to thank S.-W. Cheong, D. R. Hamann and D. Vanderbilt for valuable discussions. This work was supported by MURI--ARO Grant
W911NF-07-1-0410 and ONR Grant N0014-00-1-0302.


\begin{thebibliography}{99}
\bibitem{correction} Comparison of total energies for different supercells was made by recalculating the 5-atom 
FM phase energies with $k$-point grids compatible with the 10-atom and 20-atom supercells.
\end{thebibliography}
\end{document}